\documentclass[%
reprint,
groupedaddress,
showpacs,preprintnumbers,
amsmath,amssymb,
aps,
prl,
floatfix,
lengthcheck
]{revtex4-1}

\usepackage{graphicx}
\usepackage{dcolumn}
\usepackage{bm}
\usepackage{hyperref}

\renewcommand{\vec}[1]{\bm{#1}}

\begin{document}




\title{Collisionless plasma shocks in striated electron temperatures}


\author{P.\ Guio$^1$ and H.~L.\ P\'{e}cseli$^2$}
\affiliation{$^1$Department of Physics and Astronomy,
University College London,
Gower Street,
London WC1E 6BT,
United Kingdom\\
$^2$University of Oslo,
Department of Physics, Box 1048 Blindern, N-0316 Oslo,
Norway}
\date{\today}


\begin{abstract}
The existence of low frequency waveguide modes of ion acoustic waves is
demonstrated in magnetized plasmas for electron temperature striated along
the magnetic field lines. At higher frequencies, in a band between the ion
cyclotron and the ion plasma frequency, radiative modes develop and propagate
obliquely to the field away from the striation. Arguments for the
subsequent formation and propagation of electrostatic shock are presented
and demonstrated numerically. For such plasma conditions, the
dissipation mechanism is the ``leakage'' of the harmonics generated by the
wave steepening.
\end{abstract}

\pacs{52.35.Tc, 52.65.-y, 52.35.Fp}

\maketitle


Formation of shocks as described by Burger's equation \cite{whitham_1974}
can be understood as a balance between the energy input by an external
source (a piston moving with velocity $U$, for instance) and viscous
dissipation, with kinematic viscosity coefficient $\nu$. In one
spatial dimension, this nonlinear problem can be solved exactly
by a Cole-Hopf transformation to demonstrate, for instance, that the shock
thickness varies as 
$\sim \nu/U$ with the basic parameters of the problem. In principle,
Burger's equation can apply for any continuous viscous fluid media, also
plasmas. Experiments performed in the strongly magnetized plasma of the 
Ris{\o} Q-machine 
\cite{andersen_et_al_1967} demonstrated that for moderate electron to ion
temperature ratio $T_e/T_i$, the strong ion Landau damping prohibited the
formation 
of shock. For large temperature ratios, the ion Landau damping is reduced,
and there is a possibility for forming steady state nonlinear shock-like
forms, propagating at a constant speed
\cite{andersen_et_al_1967,ikezi_et_al_1973}. 

In the present study we present a novel mechanism of an effective energy
dissipation; 
selective radiation or ``leakage'' of short wavelength ion sound waves. We
also demonstrate 
that electrostatic shocks can form as a balance between these losses and
the standard nonlinear wave steepening as described by the nonlinear term in
the ``simple
wave'' equation, $\partial u/\partial
t+u\partial u/\partial z=0$, \cite{blackstock_1972,whitham_1974}. Studies
in two spatial dimensions are 
sufficient for illustrating the basic ideas, and  the analysis of the
present paper is restricted to 2D.

Magnetized plasmas are considered here for conditions where the electron
temperature $T_e$ varies in the direction perpendicular to an externally imposed
homogeneous magnetic field \cite{nishida_hirose_1977,guio_et_al_2001}. Such
conditions occur often in nature for plasmas out of equilibrium
\cite{penano_morales_maggs_2000}. For 
the present analysis it is essential that 
the ion cyclotron frequency is smaller than the ion plasma frequency, 
i.e.\ $\Omega_{ci}<\Omega_{pi}$. The relevant frequencies are assumed to be
so low that an inertialess electron component can be taken to be in local Boltzmann
equilibrium at all times. We assume quasi-neutrality, $n_e\approx n_i$. For
a linearized fluid model we readily derive a basic equation in the form
\begin{eqnarray}
&&\frac{\partial^4}{\partial t^4}\psi
-\frac{T_e(x)}{M}\frac{\partial^2}{\partial t^2}
\left(\nabla^2_\perp+\frac{\partial^2}{\partial z^2}\right)\psi
+\Omega_{ci}^2\frac{\partial^2}{\partial t^2}\psi\nonumber\\
&&\hspace{1cm}-\Omega_{ci}^2\frac{T_e(x)}{M}\frac{\partial^2}{\partial
  z^2}\psi = 0,
\label{basic_lineq}
\end{eqnarray}
where $\psi$ is the electrostatic potential, related to the relative
density perturbations as $e\psi/T_e=\eta\equiv n_1/n_0$. Since we are
here only interested in cases where $T_e/T_i\gg 1$, we took $T_i=0$ in
(\ref{basic_lineq}). 
In case $T_e=\mathrm{constant}$, a linear dispersion relation is readily obtained
from (\ref{basic_lineq}) by Fourier transforming with respect to time and
space. This dispersion relation contains two branches, one for
$\omega<\Omega_{ci}$ and one for $\Omega_{ci}<\omega<\Omega_{pi}$, the
latter containing also the ion cyclotron waves. The wave properties of the
two branches are very different, as illustrated best by the angle between the
group velocity and the wave-vector \cite{guio_et_al_2001}. For very low
frequencies, $\omega\ll\Omega_{ci}$, these two 
vectors are almost perpendicular, while they are close to parallel when
$\omega\gg\Omega_{ci}$. In the limit $\vec{k}_\perp\rightarrow 0$, the dispersion
relation reduces to $(\omega^2-\Omega_{ci}^2)(\omega^2-k^2C_s^2)=0$
containing ion sound waves and the electrostatic ion cyclotron resonance.

\begin{figure}
\includegraphics[width=0.99\columnwidth]{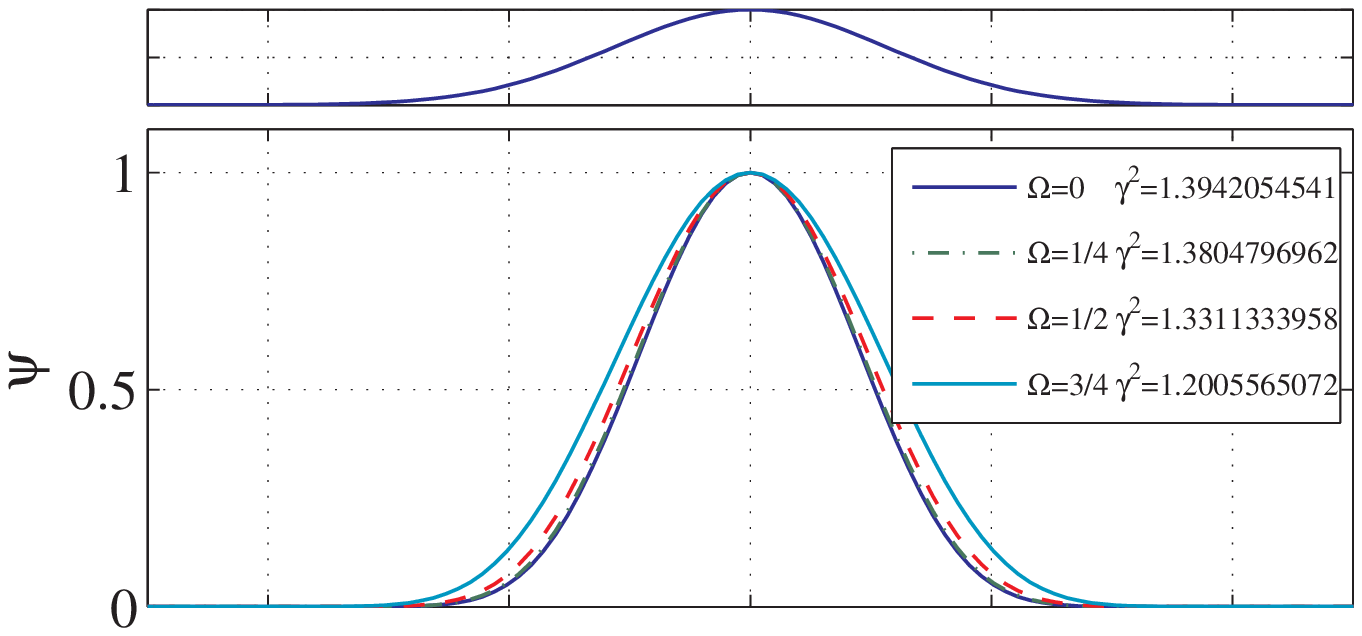}\\
\includegraphics[width=0.99\columnwidth]{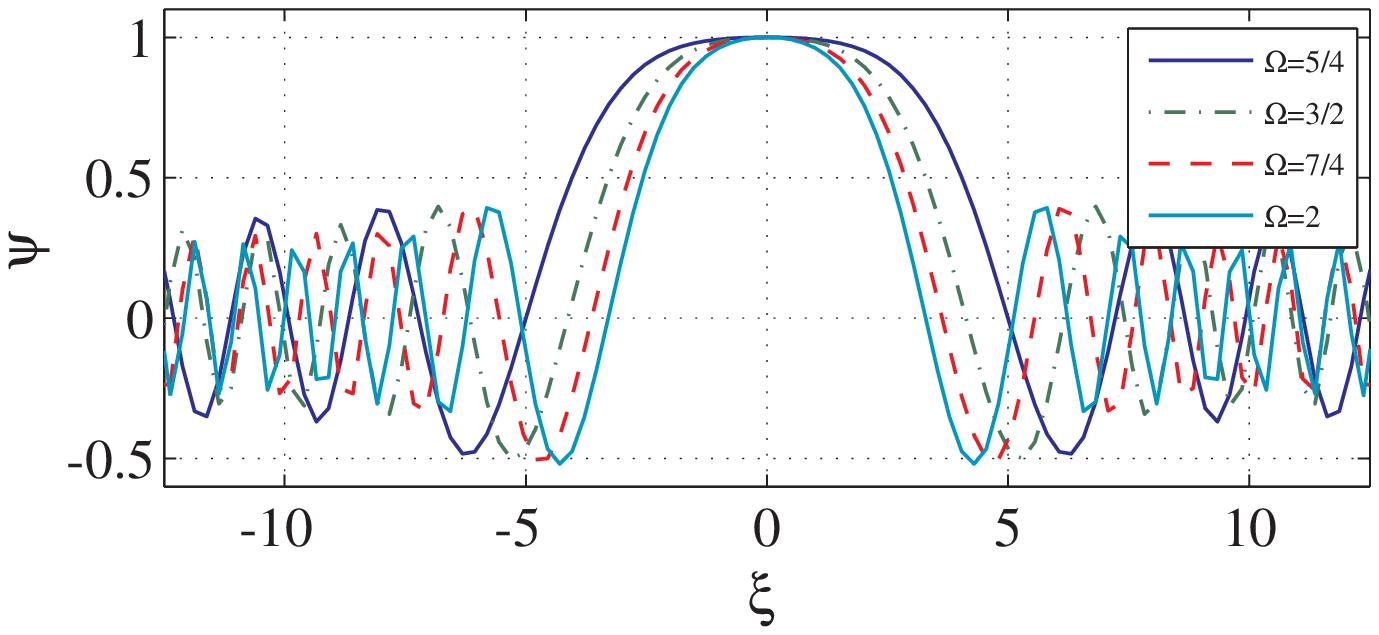}
\caption{\it Illustrative examples of numerical solutions of
  (\ref{eigenvalue_eq}) for
  the electron temperature profile shown in the upper panel, and defined by
  $T_e(\xi)/T_0=1-\frac{1}{2}{\cal D}+{\cal D}\exp(-\xi^2/{\cal W}^2)$,
  where ${\cal W}=4$ and ${\cal D}=24/23$. Waveguide mode solutions 
	for $\Omega=0,\frac{1}{4},\frac{1}{2},\frac{3}{4}$ are shown in the middle
	panel. Free modes solutions obtained for
  $\Omega=\frac{5}{4},\frac{3}{2},\frac{7}{4},2$ and $\gamma=2$ are shown in
	the lower panel.}  
\label{fig:eigenvalue}
\end{figure}

If we let $T_e=T_e(x)$, with $z$ along the
magnetic field $\vec{B}$ and $x$ in the transverse direction, we can
still Fourier transform with respect to time and the $z$-direction. We
denote the Fourier transformed electrostatic potential by $\widehat\psi$. Normalizing
frequencies and lengths so that $\Omega\equiv\omega/\Omega_{ci}$ and
$\xi\equiv x\Omega_{ci}/C_s$, respectively, we readily
obtain the expression
\begin{equation}
\frac{d^2}{d\xi^2}\widehat{\psi}=
(\Omega^2-1)\left(\frac{1}{\gamma^2}-\frac{T_0}{T_e(\xi)}\right)
\widehat{\psi},
\label{eigenvalue_eq}
\end{equation}
where we introduced a normalized propagation speed
$\gamma^2\equiv(\omega/k_z)^2(M/T_0)$, and $T_0$ is a 
reference temperature. Thus, $\gamma$ measures the ratio between the phase
velocity along the magnetic field and a sound speed, so that $\gamma=\mathrm{constant}$ would correspond to
exactly non-dispersive wave propagation. The expression
(\ref{eigenvalue_eq}) has the form of 
an eigenvalue equation, with $1/\gamma^2$ being the
eigenvalue. We present numerical
solutions in Fig.~\ref{fig:eigenvalue}. In the low frequency
limit $\omega<\Omega_{ci}$, shown in the middle panel, the waves are confined to the electron
temperature striation (here denoted ``waveguide modes''), corresponding to
a discrete set of eigenfunctions $\widehat{\psi}_m$, with mode number $m$. For
the Gaussian profile $T_e(x)$ considered here, the mode number
$m$ corresponds to the number of zero-crossing of $\widehat{\psi}_m(x)$
\cite{guio_et_al_2001}. For the  
three dimensional problem we would have {\it two} indexes $\widehat{\psi}_{km}$
corresponding to the two directions perpendicular to $\vec{B}$. From
$\widehat{\psi}_m$ we can obtain the corresponding eigenmodes for the 
$\vec{B}$-parallel velocity $u_\parallel$. The value of $\gamma$ depends on
$\Omega$, and we find for $\Omega = 0, \frac{1}{4},
\frac{1}{2}, \frac{3}{4}$, that $\gamma_0^2=
1.3942,
1.3805,
1.3311$, and
$1.2006$, i.e.\ a
relatively weak variation of $\gamma_0$ with $\Omega$. 
We note also that the eigenfunctions change only little in spite of the large
change in $\Omega$.
Only for $\Omega$ close to unity, say around 0.9 or larger, do we
see significant variations in $\widehat{\psi}_0(x)$ and $\gamma_0$.
For shallow temperature variations and narrow
temperature ducts, we have only the lowest order mode $\psi_0(x)$. For
$\gamma$ smaller 
than the minimum value of $T_e(x)/T_0$ we have a continuum of
eigenvalues with corresponding eigenfunctions. In all cases we used
$T_0\equiv \frac{1}{2}(T_e(0)+T_e(|\infty|)$.

For
$\omega>\Omega_{ci}$ the right hand side of (\ref{eigenvalue_eq}) changes
sign, and the nature of the eigenmodes changes as well, to become free
modes as seen in the lower panel in Fig.~\ref{fig:eigenvalue}. If we let the electron
temperature striation vanish to have a uniform $T_e$, then the free modes
degenerate to two obliquely propagating plane waves.

We consider now the low frequency limit of the branch of dispersion
relation with $\omega<\Omega_{ci}$. For $m>1$, the waveguide
modes can decay for one $m$-value to modes with other $m$-values. The $m=0$
mode has no decay to other forward propagating modes, and will be the
one considered here. For 
this highest phase velocity mode, with eigenmode $\widehat{\psi}_0(x)$,  
wave steepening will be the dominant nonlinearity
\cite{blackstock_1972,whitham_1974}. The nonlinear terms couple 
the various modes to give products 
of $x$-modes. These can be expanded as, for instance,
$\widehat{\psi}_{i}(x)\widehat{\psi}_{j}(x) = \sum_q 
\zeta_{qij}\widehat{\psi}_{q}(x)$, where we assume that the set $\widehat{\psi}_m$ is
complete and orthonormal \cite{manheimer_1969}. For the lowest order waveguide mode we have, in
particular,
$\zeta_{j00}=\int_{-\infty}^\infty\widehat{\psi}_{0}^2(x)\widehat{\psi}_{j}(x)dx$. For 
a large class of relevant electron temperature profiles we can ignore all 
higher 
order modes, and retain only $\widehat{\psi}_{0}$, and introduce here $\zeta_{0} 
\equiv \int_{-\infty}^\infty\widehat{\psi}_{0}^3(x)dx$. For the Gaussian
variations of $T_e(x)$ studied here, we will have $\zeta_0>\zeta_j$
for all $j\geq 1$, since $\widehat{\psi}_{0}$ is the only eigenfunction
that is positive everywhere. To lowest order in
the present low frequency limit we
have the relation $e\widehat{\psi}/T_e=\widehat{\eta}\approx
\widehat{u}_\parallel/C_{s}$ between 
fluctuations in relative 
density and the velocity in the  $\vec{B}$-parallel direction. Our
arguments concerning the mode structure therefore apply to 
the velocity variations as well.  

Considering the limit of time scales much larger than the ion cyclotron
period, we find after some algebra the result
\begin{equation}
\frac{\partial u_\parallel}{\partial t}+
(\zeta_{0} u_\parallel \pm\zeta_TC_{s0}) 
\frac{\partial u_\parallel}{\partial z}=0,
\label{simple_wave_basic}
\end{equation}
with $u_\parallel=u_\parallel(z,t)$,
where the numerical value of $\zeta_T$ is that of $\gamma_0$ in the limit
of $\Omega\rightarrow 0$. Small polarization drifts $\perp\vec{B}$ 
are ignored. We introduced a constant reference sound speed $C_{s0}$.
We anticipate that $\zeta_T$ is here not much different from $\zeta_0$, since 
the form of $\psi_{0}(x)$ is close to $T_e(x)-T_e(|\infty|)$. The solutions of
(\ref{simple_wave_basic}) 
have the well known steepening of the initial condition. The characteristic
time for wave breaking is approximately ${\cal
  L}/\mbox{max}\{u_\parallel(t=0)\}$, where ${\cal L}$ is the characteristic
scale length of the initial perturbation along $\vec{B}$ and
$\mbox{max}\{u_\parallel(t=0)\}$ is the maximum value of the initial
velocity perturbation. The model equation (\ref{simple_wave_basic}) assumes
$\widehat{\psi}_0(x)$ and $\gamma_0$ being 
used also when $\Omega>0$, but this approximation is acceptable for at
least $0\leq \Omega < 0.75$, as seen in the middle panel in Fig.~\ref{fig:eigenvalue}. For
the basic ideas outlined in the present work, this restriction is of little
consequence. Polarization drifts become increasingly larger as $\Omega$ is
increased, but these do not affect the dynamics parallel to $\vec{B}$,
which is covered by  (\ref{simple_wave_basic}).

\begin{figure}
\includegraphics[width=0.99\columnwidth]{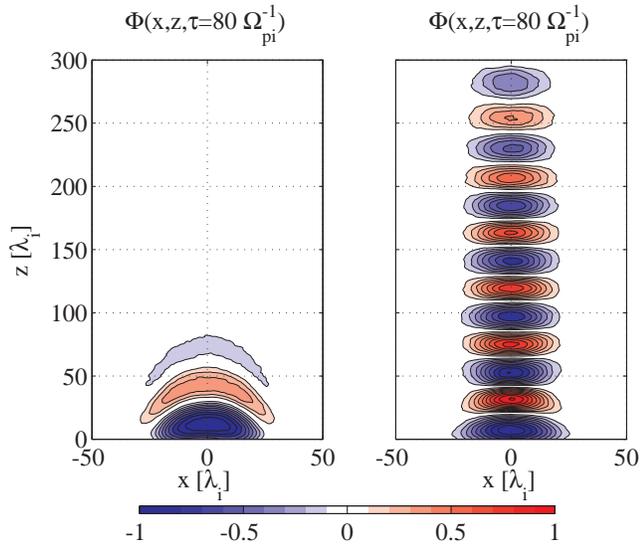}
\caption{\it Examples of numerical simulations showing the variations of
  the normalized potential $\Phi=e\psi/T_i$ as a function of position at a fixed time. The electron
  temperature enhancement is localized as a 
  Gaussian in the $x$-direction. We have $\omega=\Omega_{pi}\pi/5$ in both
  cases while $\Omega_{ci}/\Omega_{pi}=0.05$ and
  $\Omega_{ci}/\Omega_{pi}=1$ in the left and right panels respectively. We
  have here $T_e/T_i=50$. For clarity, only a part of the simulation domain
  is shown.} 
\label{fig:numres_modes}
\end{figure}

If we initialize the system with characteristic wavelengths corresponding to
frequencies $\omega\ll\Omega_{ci}$, i.e.\ ${\cal L}\gg C_s/\Omega_{ci}$, the
short time evolution will be governed 
by (\ref{simple_wave_basic}), and we will have shorter and shorter scales
developing as for the usual breaking of waves
\cite{blackstock_1972,whitham_1974}. This process is however 
arrested when the characteristic length scales become of the order of
the effective ion Larmor radius $C_s/\Omega_{ci}$, where the modes become
radiating, and are no longer confined 
to the waveguide. We propose a phenomenological expression for the
process, best written in Fourier space in a frame moving with $C_{s0}$, as
\begin{equation}
\frac{\partial\widehat{u}_\parallel}{\partial t} + i\,
\frac{\zeta_{0} k_\parallel}{2} \widehat{u}_\parallel \otimes
\widehat{u}_\parallel =
-\frac{\widehat{u}_\parallel}{\mathop{\cal T}(k_\parallel)}
\mathop{\cal H}\left(|k_\parallel|-\frac{\Omega_{ci}}{C_{s0}}\right),
\label{simple_model}
\end{equation}
with $\widehat{u}_\parallel = \widehat{u}_\parallel(k_\parallel,t)$.
The symbol $\otimes$ denotes the convolution product and ${\cal H}$ is Heaviside's
step function. ${\cal T}$ characterizes the time
it takes for the energy of the $\omega > \Omega_{ci}$-waves to be lost from
the waveguide. We have ${\cal T}={\cal T}(k_\parallel)$, but it
will depend
also on parameters such as the waveguide width as well as the other
plasma parameters. We expect that increasing width gives increasing ${\cal
  T}$, i.e.\ decreasing shock thickness $\Delta$.
Within the present model the waveform will
steepen uninhibited until 
the shock thickness becomes of the order of $C_s/\Omega_{ci}$, at
which time the harmonic frequencies will exceed $\Omega_{ci}$ to become
radiative and the high frequency wave energy is lost from the
waveguide. Consequently, the $\vec{B}$-parallel scale of the shock is
controlled by a quantity referring to the $\vec{B}$-perpendicular
dynamics.

\begin{figure}
\includegraphics[width=0.99\columnwidth]{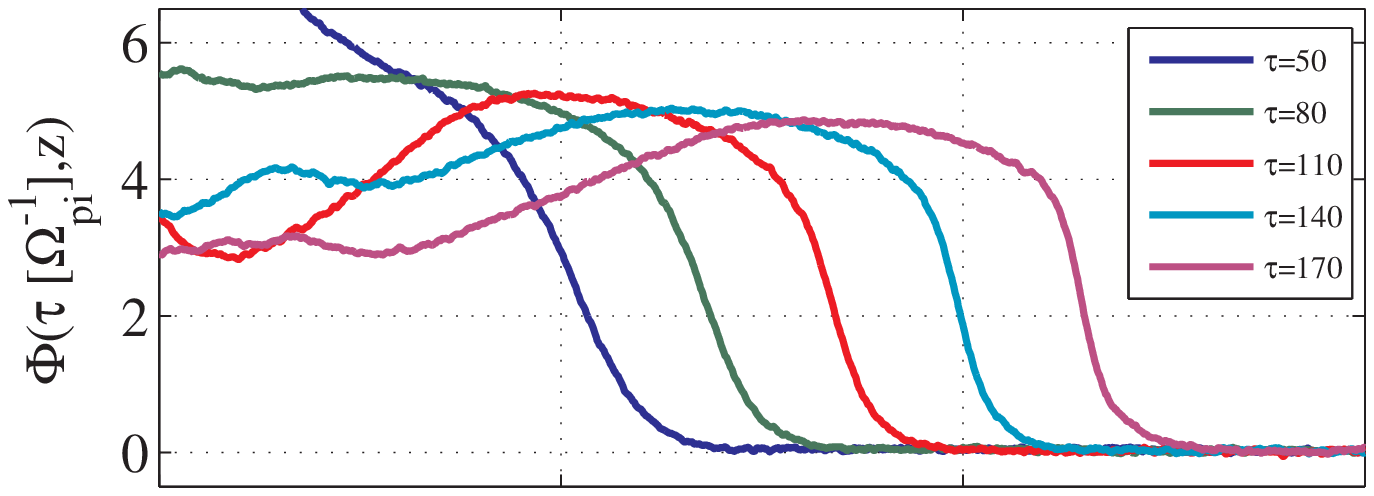}
\includegraphics[width=0.99\columnwidth]{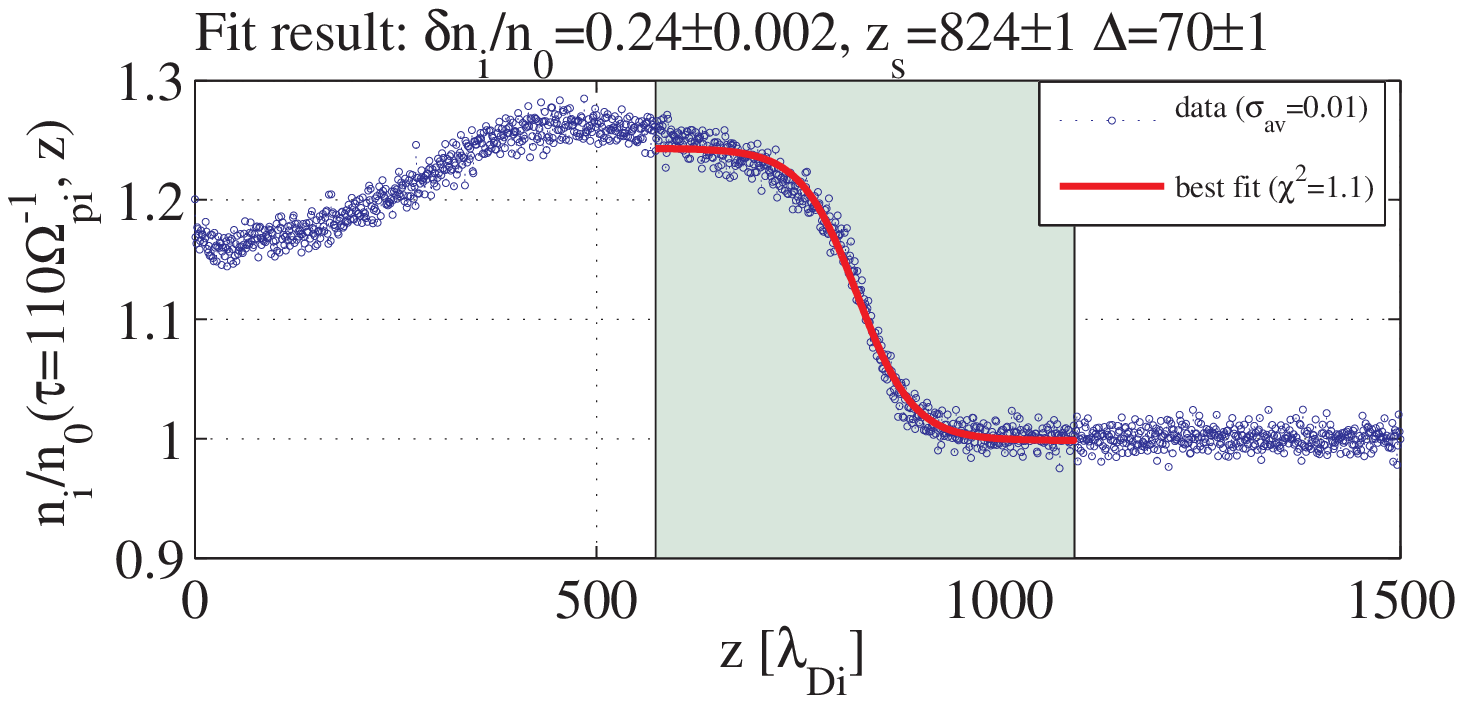}
\caption{\it Numerical simulation showing normalized potential
  $\Phi=e\psi/T_i$  and relative density 
  $n_i/n_0$ during the formation and propagation of a shock under the
  conditions mentioned before. We have $T_e/T_i=25$,
  $\Omega_{ci}/\Omega_{pi}=\frac{1}{2}$ and $\delta n_i/n_0 \approx 0.24$, where
  $\delta n_i$ is the actual detected density perturbation at the time
  where the shock is fully formed. The background density $n_0$ is
  normalized to unity. Lower frame shows sample of shock fitting for the
  ion density.} 
\label{fig:numres_shock}
\end{figure}

We study the nonlinear propagation of low frequency waves in an electron
temperature striated magnetized 
plasma numerically by using a $2\frac{1}{2}$-dimensional particle in cell
code described 
elsewhere \cite{guio_et_al_2001}. The code assumes explicitly the
electrons to be locally Boltzmann distributed and the resulting nonlinear
Poisson equation is solved by iteration. We use a Gaussian variation for 
$T_e(x)$ so that $T_e(|\infty|)/T_i=1$, while
$T_e(0)/T_i>1$, where $T_i=\mathrm{constant}$. The width of the electron
striation is 
here ${\cal W}= 28 \,\lambda_{Di}=3.9 \,C_{s0}/\Omega_{ci}$. Several values
of $T_e(0)/T_i$ were investigated.

Results illustrating the waveguide and
the free modes are shown in Fig.~\ref{fig:numres_modes}. The properties
are clearly different and consistent with the interpretation given before.
The waveguide modes are confined to the electron striation, while the high
frequency free modes are dispersing or ``leaking'', consistent also with
laboratory experimental 
results \cite{nishida_hirose_1977}. It is important to emphasize that this
apparent damping will be found also in a fluid model. The observed
effective damping is caused by wave energy dispersing in space, and
dissipated by linear 
ion Landau damping outside the electron temperature striations where
$T_e(|\infty|)/T_i=1$ in all cases considered. Each step in the
process is formally  time-reversible. In order to  
emphasize the physical effects we discuss here, we consider only high
temperature ratios, $T_e(0)/T_i \geq 25$, in order to reduce the effects of linear
as well as nonlinear ion Landau damping. Such high temperature
ratios (even as large as $T_e/T_i=100$) can be obtained in discharge plasmas under laboratory
conditions \cite{takahashi_et_al_1998}. For nonlinear waves described by a
Korteweg-de Vries equation, a shock-like structure is followed by Airy-type
ripples, originating from the dispersion term in the KdV-equation. These
ripples are absent in our results, see
Fig.~\ref{fig:numres_shock}. Likewise, at the high temperature ratio used
here (with $C_{s0}\gg u_{th,i}$, the ion thermal velocity), we find no ions
being reflected by the shock. A backward propagating 
rarefaction wave is of no concern here.

\begin{figure}
\includegraphics[width=0.99\columnwidth]{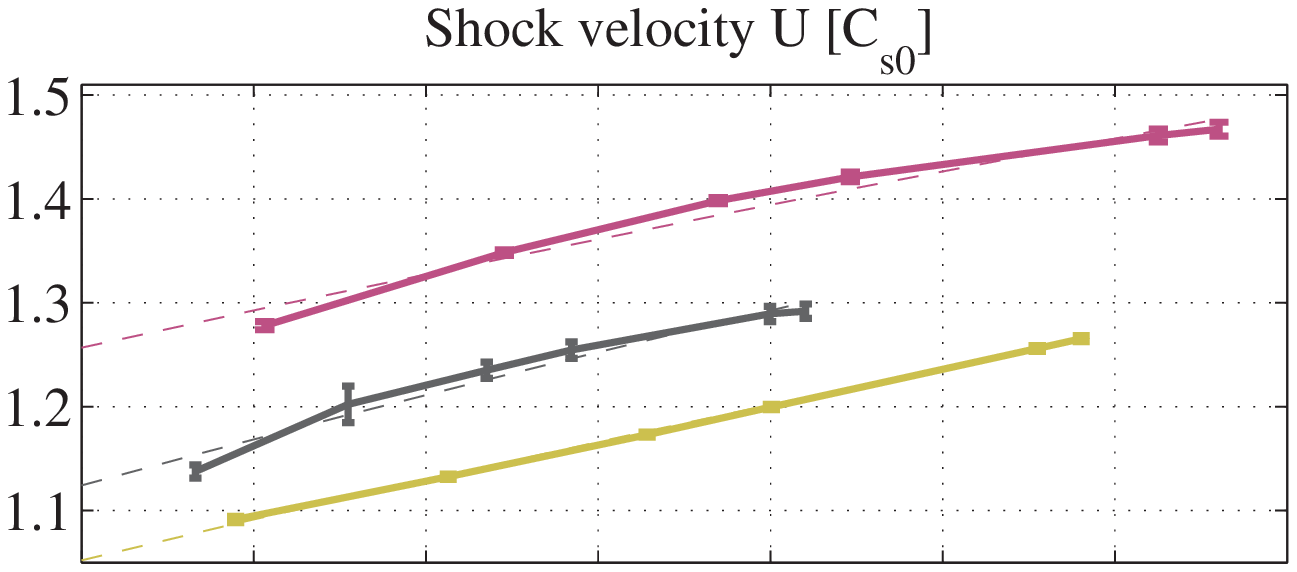}\\
\includegraphics[width=0.99\columnwidth]{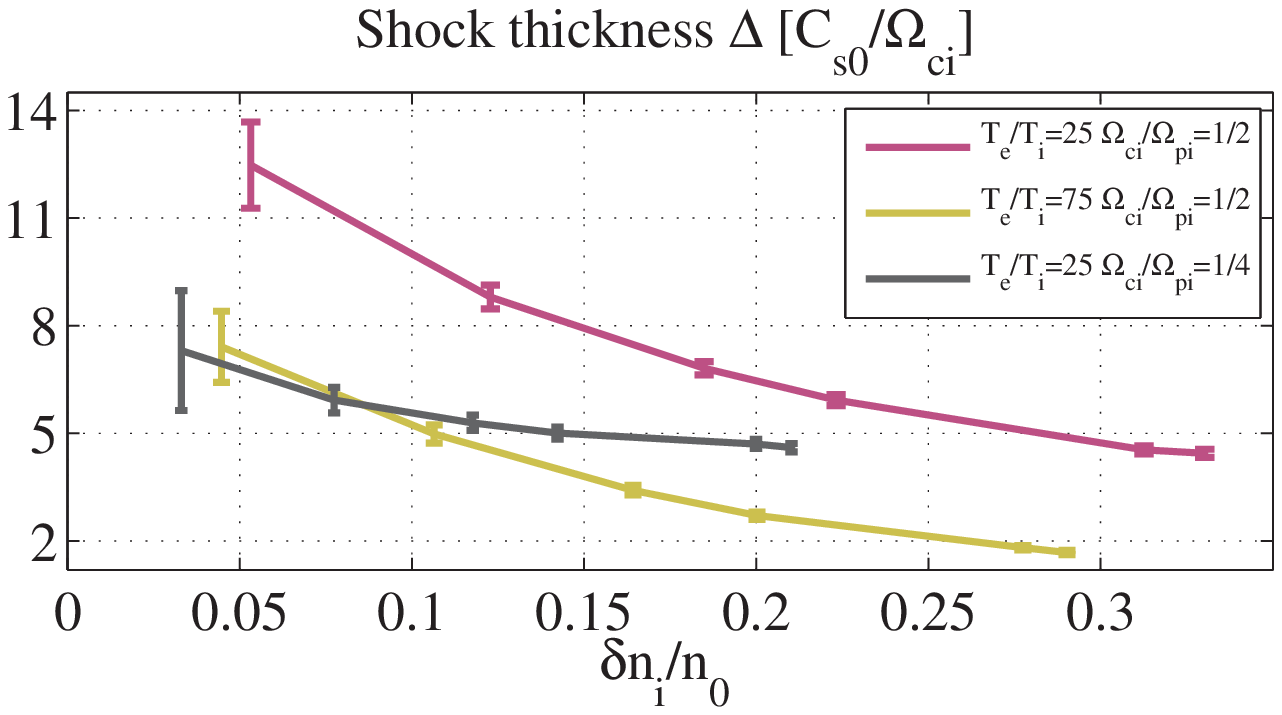}
\caption{\it Shock velocity $U$ in units of $C_{s0}$ 
  (upper panel) and saturated shock thickness $\Delta$ in units of
	$C_{s0}/\Omega_{ci}$ (lower panel) as function of 
	the fully formed shock relative height  $\delta n_i/n_0$, 
  for different combinations of the parameters $T_e/T_i=25$ and 75, and
  $\Omega_{ci}/\Omega_{pi}=\frac{1}{4}$ and $\frac{1}{2}$.}
\label{fig:numres_shock_data}
\end{figure}

In Fig.~\ref{fig:numres_shock} we show an example of the
formation and propagation of a shock. The initial density has an error
function type spatial  
variation in the $\vec{B}$-field aligned direction. Ions are continuously
injected at the boundary at $z=0$ to maintain the 
energy input. We have $T_e/T_i=25$,
$\Omega_{ci}/\Omega_{pi}=\frac{1}{2}$ and $\delta 
n_i/n_0 \approx 0.21$, where $\delta n_i$ refers to the actual detected density
perturbation at the time where the shock is fully formed, and not to the
initial imposed perturbation. 
The standard deviation of the parameters is estimated from fitting the
shock profile with a nonlinear Levenberg-Marquardt method.

In Fig.~\ref{fig:numres_shock_data} we show the shock thickness $\Delta$
and the normalized shock velocity $U$ for various combinations of the parameters
$T_e/T_i$, $\Omega_{ci}/\Omega_{pi}$ and $\delta n_i/n_0$. The velocity $U$
is obtained by following the shock position $z_s$ as function of time. For small
or moderate values of $\delta n_i/n_0$ we find a close to linear relationship
between $\delta n_i$ and $U$. As $\delta n_i\rightarrow 0$ (shown with
dashed line fit) we have $U$
approaching the appropriate value for $\gamma_0$, apart from small
corrections due to finite $T_i$ ignored in (\ref{eigenvalue_eq}). For
$\Omega_{ci}/\Omega_{pi}=\frac{1}{2}$ we find that an 
approximate five-fold increase in 
$\delta n_i/n_0$ corresponds to approximately $50\,\%$ reduction in
$\Delta$. We demonstrated also that
the shock thickness $\Delta$ scales inversely proportional to the width of the
striation. We verified that changes in the initial $\vec{B}$-parallel
scale length of the initial 
density profile do not change the saturated shock thickness $\Delta$ for any $\delta 
n_i/n_0$. We studied also a weak magnetic field limit, with
$\Omega_{ci}/\Omega_{pi} = \frac{1}{4}$. Here, finite ion Larmor radius effects
(ignored in deriving (\ref{basic_lineq})) begin to be important, since the
striation is now only $\sim 10$ ion Larmor radii wide. For this limit we
find that $\Delta$ is almost constant. Even weaker magnetic fields will
require a fully kinetic theoretical analysis to account for the effects of
collisionless ion viscosity \cite{braginskii_1965}.

We here reported arguments for the formation of
shocks in electron striations in magnetized plasmas when  $\Omega_{ci} <
\Omega_{pi}$ and $T_e\gg T_i$. By PIC-simulations we demonstrated the
formation and 
propagation of electrostatic shocks under these conditions for a wide
combination of parameters. The
decreasing shock thickness for increasing amplitudes is found also for
classical shocks as described by Burger's equation, but in our case the
dissipation mechanism is leakage from the electron temperature striation of
the short scale 
lengths generated by the nonlinear wave steepening. The ideas presented
here can have wider applications.


\begin{thebibliography}{10}%
\makeatletter
\providecommand \@ifxundefined [1]{%
 \ifx #1\undefined \expandafter \@firstoftwo
 \else \expandafter \@secondoftwo
\fi
}%
\providecommand \@ifnum [1]{%
 \ifnum #1\expandafter \@firstoftwo
 \else \expandafter \@secondoftwo
\fi
}%
\providecommand \enquote [1]{``#1''}%
\providecommand \bibnamefont  [1]{#1}%
\providecommand \bibfnamefont [1]{#1}%
\providecommand \citenamefont [1]{#1}%
\providecommand\href[0]{\@sanitize\@href}%
\providecommand\@href[1]{\endgroup\@@startlink{#1}\endgroup\@@href}%
\providecommand\@@href[1]{#1\@@endlink}%
\providecommand \@sanitize [0]{\begingroup\catcode`\&12\catcode`\#12\relax}%
\@ifxundefined \pdfoutput {\@firstoftwo}{%
 \@ifnum{\z@=\pdfoutput}{\@firstoftwo}{\@secondoftwo}%
}{%
 \providecommand\@@startlink[1]{\leavevmode}%
 \providecommand\@@endlink[0]{}%
}{%
 \providecommand\@@startlink[1]{%
  \leavevmode
  \pdfstartlink
   attr{/Border[0 0 1 ]/H/I/C[0 1 1]}%
   user{/Subtype/Link/A<</Type/Action/S/URI/URI(#1)>>}%
  \relax
 }%
 \providecommand\@@endlink[0]{\pdfendlink}%
}%
\providecommand \url  [0]{\begingroup\@sanitize \@url }%
\providecommand \@url [1]{\endgroup\@href {#1}{\urlprefix}}%
\providecommand \urlprefix [0]{URL }%
\providecommand \Eprint[0]{\href }%
\@ifxundefined \urlstyle {%
  \providecommand \doi [1]{doi:\discretionary{}{}{}#1}%
}{%
  \providecommand \doi [0]{doi:\discretionary{}{}{}\begingroup
  \urlstyle{rm}\Url }%
}%
\providecommand \doibase [0]{http://dx.doi.org/}%
\providecommand \Doi[1]{\href{\doibase#1}}%
\providecommand \bibAnnote [3]{%
  \BibitemShut{#1}%
  \begin{quotation}\noindent
    \textsc{Key:}\ #2\\\textsc{Annotation:}\ #3%
  \end{quotation}%
}%
\providecommand \bibAnnoteFile [2]{%
  \IfFileExists{#2}{\bibAnnote {#1} {#2} {\input{#2}}}{}%
}%
\providecommand \typeout [0]{\immediate \write \m@ne }%
\providecommand \selectlanguage [0]{\@gobble}%
\providecommand \bibinfo [0]{\@secondoftwo}%
\providecommand \bibfield [0]{\@secondoftwo}%
\providecommand \translation [1]{[#1]}%
\providecommand \BibitemOpen[0]{}%
\providecommand \bibitemStop [0]{}%
\providecommand \bibitemNoStop [0]{.\EOS\space}%
\providecommand \EOS [0]{\spacefactor3000\relax}%
\providecommand \BibitemShut [1]{\csname bibitem#1\endcsname}%
\bibitem{whitham_1974}%
  \BibitemOpen
  \bibfield{author}{%
  \bibinfo {author} {\bibfnamefont{G.~B.}\ \bibnamefont{Whitham}},\ }%
  \emph{\bibinfo {title} {Linear and Nonlinear Waves}}\ (\bibinfo {publisher}
  {John Wiley \& Sons},\ \bibinfo {address} {New York},\ \bibinfo {year}
  {1974})%
  \bibAnnoteFile{NoStop}{whitham_1974}%
\bibitem{andersen_et_al_1967}%
  \BibitemOpen
  \bibfield{author}{%
  \bibinfo {author} {\bibfnamefont{H.~K.}\ \bibnamefont{{Andersen}}}, \bibinfo
  {author} {\bibfnamefont{N.}~\bibnamefont{{D'Angelo}}}, \bibinfo {author}
  {\bibfnamefont{P.}~\bibnamefont{{Michelsen}}},\ and\ \bibinfo {author}
  {\bibfnamefont{P.}~\bibnamefont{{Nielsen}}},\ }%
  \bibfield{journal}{%
  \Doi{10.1103/PhysRevLett.19.149}{\bibinfo {journal} {Phys. Rev. Lett.}}\ }%
  \textbf{\bibinfo {volume} {19}},\ \bibinfo {pages} {149} (\bibinfo {month}
  {Jul.}\ \bibinfo {year} {1967})%
  \bibAnnoteFile{NoStop}{andersen_et_al_1967}%
\bibitem{ikezi_et_al_1973}%
  \BibitemOpen
  \bibfield{author}{%
  \bibinfo {author} {\bibfnamefont{H.}~\bibnamefont{{Ikezi}}}, \bibinfo
  {author} {\bibfnamefont{T.}~\bibnamefont{{Kamimura}}}, \bibinfo {author}
  {\bibfnamefont{M.}~\bibnamefont{{Kako}}},\ and\ \bibinfo {author}
  {\bibfnamefont{K.~E.}\ \bibnamefont{{Lonngren}}},\ }%
  \bibfield{journal}{%
  \Doi{10.1063/1.1694282}{\bibinfo {journal} {Phys. Fluids}}\ }%
  \textbf{\bibinfo {volume} {16}},\ \bibinfo {pages} {2167} (\bibinfo {month}
  {Dec.}\ \bibinfo {year} {1973})%
  \bibAnnoteFile{NoStop}{ikezi_et_al_1973}%
\bibitem{blackstock_1972}%
  \BibitemOpen
  \bibfield{author}{%
  \bibinfo {author} {\bibfnamefont{D.~T.}\ \bibnamefont{Blackstock}},\ }%
  \enquote{\bibinfo {title} {Nonlinear acoustics (theoretical)},}\ in\
  \emph{\bibinfo {booktitle} {American Institute of Physics Handbook}}\
  (\bibinfo {publisher} {McGraw-Hill},\ \bibinfo {address} {New York},\
  \bibinfo {year} {1972})\ pp.\ \bibinfo {pages} {3--183},\ \bibinfo {edition}
  {3rd}\ ed.%
  \bibAnnoteFile{Stop}{blackstock_1972}%
\bibitem{nishida_hirose_1977}%
  \BibitemOpen
  \bibfield{author}{%
  \bibinfo {author} {\bibfnamefont{Y.}~\bibnamefont{{Nishida}}}\ and\ \bibinfo
  {author} {\bibfnamefont{A.}~\bibnamefont{{Hirose}}},\ }%
  \bibfield{journal}{%
  \Doi{10.1088/0032-1028/19/5/005}{\bibinfo {journal} {Phys. Plasmas}}\ }%
  \textbf{\bibinfo {volume} {19}},\ \bibinfo {pages} {447} (\bibinfo {month}
  {May}\ \bibinfo {year} {1977})%
  \bibAnnoteFile{NoStop}{nishida_hirose_1977}%
\bibitem{guio_et_al_2001}%
  \BibitemOpen
  \bibfield{author}{%
  \bibinfo {author} {\bibfnamefont{P.}~\bibnamefont{Guio}}, \bibinfo {author}
  {\bibfnamefont{S.}~\bibnamefont{B{\o}rve}}, \bibinfo {author}
  {\bibfnamefont{H.~L.}\ \bibnamefont{P{\'e}cseli}},\ and\ \bibinfo {author}
  {\bibfnamefont{J.}~\bibnamefont{Trulsen}},\ }%
  \bibfield{journal}{%
  \bibinfo {journal} {Ann. Geophysicae}\ }%
  \textbf{\bibinfo {volume} {18}},\ \bibinfo {pages} {1613} (\bibinfo {year}
  {2001})%
  \bibAnnoteFile{NoStop}{guio_et_al_2001}%
\bibitem{penano_morales_maggs_2000}%
  \BibitemOpen
  \bibfield{author}{%
  \bibinfo {author} {\bibfnamefont{J.~R.}\ \bibnamefont{Pe{\~n}ano}}, \bibinfo
  {author} {\bibfnamefont{G.~J.}\ \bibnamefont{Morales}},\ and\ \bibinfo
  {author} {\bibfnamefont{J.~E.}\ \bibnamefont{Maggs}},\ }%
  \bibfield{journal}{%
  \bibinfo {journal} {Phys. Plasmas}\ }%
  \textbf{\bibinfo {volume} {7}},\ \bibinfo {pages} {144} (\bibinfo {year}
  {2000})%
  \bibAnnoteFile{NoStop}{penano_morales_maggs_2000}%
\bibitem{manheimer_1969}%
  \BibitemOpen
  \bibfield{author}{%
  \bibinfo {author} {\bibfnamefont{W.~M.}\ \bibnamefont{{Manheimer}}},\ }%
  \bibfield{journal}{%
  \Doi{10.1063/1.1692362}{\bibinfo {journal} {Phys. Fluids}}\ }%
  \textbf{\bibinfo {volume} {12}},\ \bibinfo {pages} {2426} (\bibinfo {month}
  {Nov.}\ \bibinfo {year} {1969})%
  \bibAnnoteFile{NoStop}{manheimer_1969}%
\bibitem{takahashi_et_al_1998}%
  \BibitemOpen
  \bibfield{author}{%
  \bibinfo {author} {\bibfnamefont{K.}~\bibnamefont{{Takahashi}}}, \bibinfo
  {author} {\bibfnamefont{T.}~\bibnamefont{{Oishi}}}, \bibinfo {author}
  {\bibfnamefont{K.}~\bibnamefont{{Shimomai}}}, \bibinfo {author}
  {\bibfnamefont{Y.}~\bibnamefont{{Hayashi}}},\ and\ \bibinfo {author}
  {\bibfnamefont{S.}~\bibnamefont{{Nishino}}},\ }%
  \bibfield{journal}{%
  \Doi{10.1103/PhysRevE.58.7805}{\bibinfo {journal} {Phys. Rev. E}}\ }%
  \textbf{\bibinfo {volume} {58}},\ \bibinfo {pages} {7805} (\bibinfo {month}
  {Dec.}\ \bibinfo {year} {1998})%
  \bibAnnoteFile{NoStop}{takahashi_et_al_1998}%
\bibitem{braginskii_1965}%
  \BibitemOpen
  \bibfield{author}{%
  \bibinfo {author} {\bibfnamefont{S.~I.}\ \bibnamefont{Braginski\'{\i}}},\ }%
  \enquote{\bibinfo {title} {Transport processes in a plasma},}\ in\
  \emph{\bibinfo {booktitle} {Reviews of plasma physics}},\ Vol.~\bibinfo
  {volume} {1},\ \bibinfo {editor} {edited by\ \bibinfo {editor}
  {\bibfnamefont{M.~A.}\ \bibnamefont{Leontovich}}}\ (\bibinfo {publisher}
  {Consultants Bureau, New York},\ \bibinfo {year} {1965})\ pp.\ \bibinfo
  {pages} {205--311}%
  \bibAnnoteFile{NoStop}{braginskii_1965}%
\end{thebibliography}
\end{document}